\documentclass[a4paper]{jpconf}
\usepackage{graphicx}
\usepackage{amsmath,amssymb,xspace}
\usepackage[breaklinks]{hyperref}

\ifdefined\draftflag
\usepackage{lineno}
\linenumbers
\fi

\newcommand{\PYTHIA} {{\textsc{pythia}}\xspace}
\newcommand{\GEANTfour} {{\textsc{Geant4}}\xspace}
\newcommand{\unit}[1]{\ensuremath{\text{\,#1}}\xspace}
\newcommand{\MeV}{\ensuremath{\,\text{Me\hspace{-.08em}V}}\xspace}
\newcommand{\GeV}{\ensuremath{\,\text{Ge\hspace{-.08em}V}}\xspace}
\newcommand{\Eprime}{\ensuremath{E^{\prime}}\xspace}
\newcommand{\Enorm}{\ensuremath{E_{\text{norm}}}\xspace}
\newcommand{\Epixel}{\ensuremath{E_{\text{pixel}}}\xspace}
\newcommand{\concordance}{\ensuremath{\rho_{c}}\xspace}

\newlength{\myfigurewidth}
\newlength{\myfigureskip}
\newcommand{\threefig}[3]{\includegraphics[width=0.33\myfigurewidth]{#1}\hspace{\myfigureskip}\includegraphics[width=0.33\myfigurewidth]{#2}\hspace{\myfigureskip}\includegraphics[width=0.33\myfigurewidth]{#3}}
\newcommand{\twofigthree}[2]{\includegraphics[width=0.33\myfigurewidth]{#1}\hspace{\myfigureskip}\includegraphics[width=0.33\myfigurewidth]{#2}}
\newlength{\myfigureskipeqh}
\newcommand{\twofigeqh}[2]{\resizebox{\myfigurewidth}{!}{\includegraphics[height=1cm]{#1}\hspace{\myfigureskipeqh}\includegraphics[height=1cm]{#2}}}

\begin{document}

\title{Denoising Convolutional Networks to Accelerate Detector Simulation}
\ifdefined\draftflag
\thispagestyle{plain}
\fi

\author{S Banerjee$^1$, B Cruz Rodriguez$^2$, L Franklin$^3$, H Guerrero De La Cruz$^2$, T Leininger$^4$,
S Norberg$^2$, K Pedro$^5$, A Rosado Trinidad$^2$, and Y Ye$^6$ (on behalf of the CMS Collaboration)}

\address{$^1$ University of Wisconsin, Madison}
\address{$^2$ University of Puerto Rico, Mayaguez}
\address{$^3$ University of Maryland, College Park}
\address{$^4$ Lafayette University}
\address{$^5$ Fermi National Accelerator Laboratory}
\address{$^6$ University of Chicago}

\ead{pedrok@cern.ch}

\begin{abstract}
The high accuracy of detector simulation is crucial for modern particle physics experiments.
However, this accuracy comes with a high computational cost,
which will be exacerbated by the large datasets and complex detector upgrades
associated with next-generation facilities such as the High Luminosity LHC.
We explore the viability of regression-based machine learning (ML) approaches using convolutional neural networks (CNNs) to ``denoise'' faster, lower-quality detector simulations,
augmenting them to produce a higher-quality final result with a reduced computational burden.
The denoising CNN works in concert with classical detector simulation software rather than replacing it entirely,
increasing its reliability compared to other ML approaches to simulation.
We obtain promising results from a prototype based on photon showers in the CMS electromagnetic calorimeter.
Future directions are also discussed. 
\end{abstract}

\section{Introduction}

Accurate detector simulation is one of the cornerstones of particle physics, crucial for everything from detector design to data analysis.
The CMS detector~\cite{Chatrchyan:2008zzk} is simulated with \GEANTfour~\cite{Agostinelli:2002hh,Allison:2016lfl}, which comes with a substantial computational burden:
it consumed 40\% of grid CPU at the beginning of the LHC Run 2~\cite{Apostolakis:2018ieg}.
The CMS detector simulation already benefits from numerous technical optimizations and physics-preserving approximations,
which improve its throughput by a factor of 4--6 compared to the default~\cite{Pedro:2019mkq}.
After the HL-LHC upgrade, the CPU time needed to simulate an event will increase by a factor of 3 or more~\cite{Pedro:2020kbk},
because the accompanying detector upgrades introduce a more complex geometry and require more
detailed physics models to reproduce their precise measurements.
Machine learning (ML) is a promising avenue to address this challenge by producing high-quality simulated events more quickly.

We investigate the performance of ``denoising'', a regression-based ML approach
that enhances faster, lower-quality detector simulations to produce a higher-quality final result with a reduced computational burden.
This approach is based on Monte Carlo (MC) ray-tracing used in computer animation~\cite{Pixar}.
Because a classical simulation engine still simulates every event, reliability concerns from attempting to generate wholly novel output,
e.g. with a generative adversarial network (GAN), are naturally avoided.
In particular, we use a version of \GEANTfour with modified parameter settings as the faster, lower-quality simulation engine.
As a proof of concept, we train the denoising neural network (NN) on photon showers in the CMS electromagnetic calorimeter (ECAL)
and achieve results competitive with other ML approaches.
The results in this proceeding are the first application of denoising to MC detector simulation.
A similar approach called ``super-resolution'' was recently studied to improve jet reconstruction~\cite{Baldi:2020hjm}.

\section{Simulation}

In order to adapt \GEANTfour to be used as a faster simulation engine,
we study the impact of different parameter modifications on the time to simulate minimum bias events.
These studies were conducted with \GEANTfour 10.4.3~\cite{Geant4SW} and CMS software version CMSSW\_11\_0\_0~\cite{CMSSW}.

\begin{figure}[htb!]
\centering
\includegraphics[width=0.49\myfigurewidth]{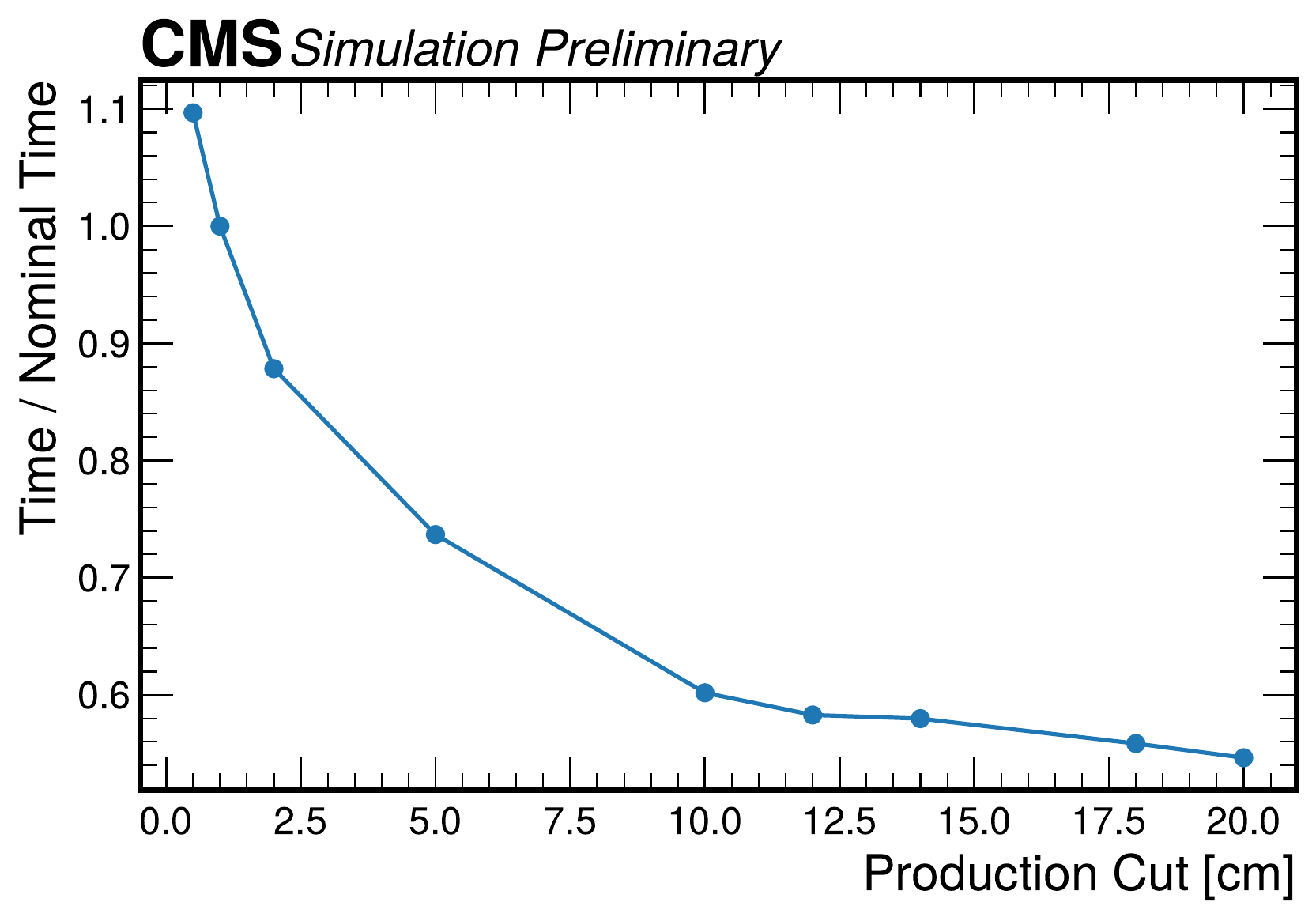}
\caption{The relative computing time to process 1000 minimum bias events in \GEANTfour for different values of the production cut (in cm).
}
\label{fig:prodcut}
\end{figure}

The most pronounced impact is obtained by varying the production cut.
A particle that would deposit all of its energy in a distance less than the production cut
will not produce any secondary particles, reducing the overall time to simulate its interactions with the detector.
Figure~\ref{fig:prodcut} shows that increasing the production cut decreases the time per event by almost a factor of two
compared to the default value of 1\unit{cm}.
For the purposes of this study, we define the faster, modified simulation to use a production cut of 10\unit{cm}.

Several other parameters are investigated and found to have minor impacts.
The CMS simulation already employs an algorithm called Russian Roulette~\cite{Lange:2015sba},
which discards $N-1$ randomly-selected low-energy particles, retaining the $N^{\text{th}}$ particle
and multiplying its energy by $N$ to account for the energies of the discarded particles without actually simulating their interactions.
The maximum energy threshold for particles considered by this algorithm can be specified independently for each type: electron, neutron, photon, or proton.
We find that increasing the neutron or photon thresholds from their default values of 10\MeV and 5\MeV, respectively, up to 200\MeV
decreases the simulation time by 4.4\% and 6.5\%, with diminishing returns from further increases.
There is also an energy threshold below which the simulation uses a simplified magnetic field propagation algorithm based on chord finding,
rather than the default, more computationally intensive G4TDormandPrince45 algorithm.
Increasing this threshold from the default 15\MeV to 350\MeV decreases the simulation time by 7.3\%.
Because the impacts of these changes are not substantial, they are not included in the modified simulation used in subsequent sections.
More parameters remain to be evaluated.

\section{Training}

We generate single photon showers using the \PYTHIA~\cite{Sjostrand:2014zea} generator (version 8.243) with energy $E = 850\GeV$ and $\eta = 0.5$, $\phi = 0.0$, corresponding to the ECAL barrel.
High-energy photons are chosen because their showers have less variation than lower-energy photons,
so a smaller training dataset suffices for the NN to learn their behavior.
The \GEANTfour energy deposits are discretized into a $50\times50$ grid in the $x$ and $y$ coordinates,
with the shower energy summed over the $z$ coordinate as a simplification to produce two-dimensional images.
The same generated events are processed with the high-quality CMS simulation and the modified simulation with the production cut set to 10\unit{cm}~\cite{SimDenoising}.
The simulation random seed is set to a specified, independent value for each event for consistency between the two simulations.
Three datasets of 5000 events are produced for training, validation, and testing.

The training data are pre-processed before being input to the NN.
The images are randomly rotated and flipped to prevent the NN from learning a preferred orientation.
The shower images have one feature per pixel, the deposited energy, which is transformed into $\Eprime = \sqrt{E}$.
This is found to provide better low-energy fidelity than similar transforms such as $\log{E}$.
The transformed energy is then normalized as $\Enorm = (\Eprime - \mu)/\sigma$,
where $\mu$ and $\sigma$ are the mean and standard deviation of the \Eprime values in all pixels,
computed per-event (image) from the modified simulation.

\begin{figure}[htb!]
\centering
\twofigeqh{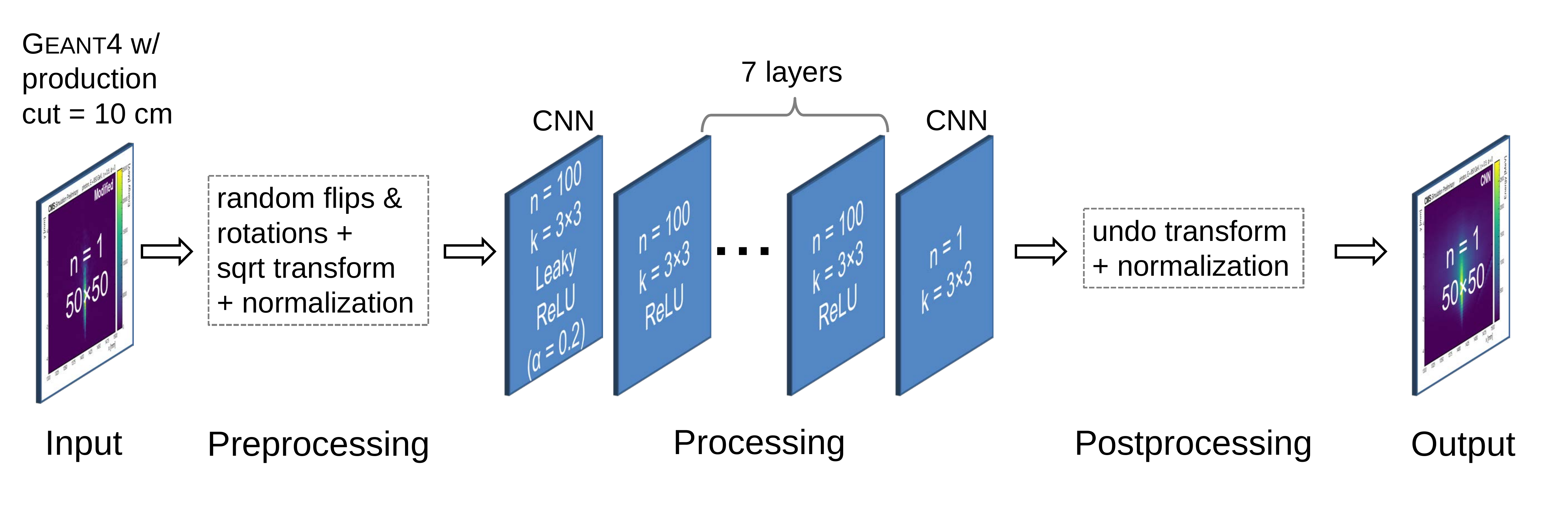}{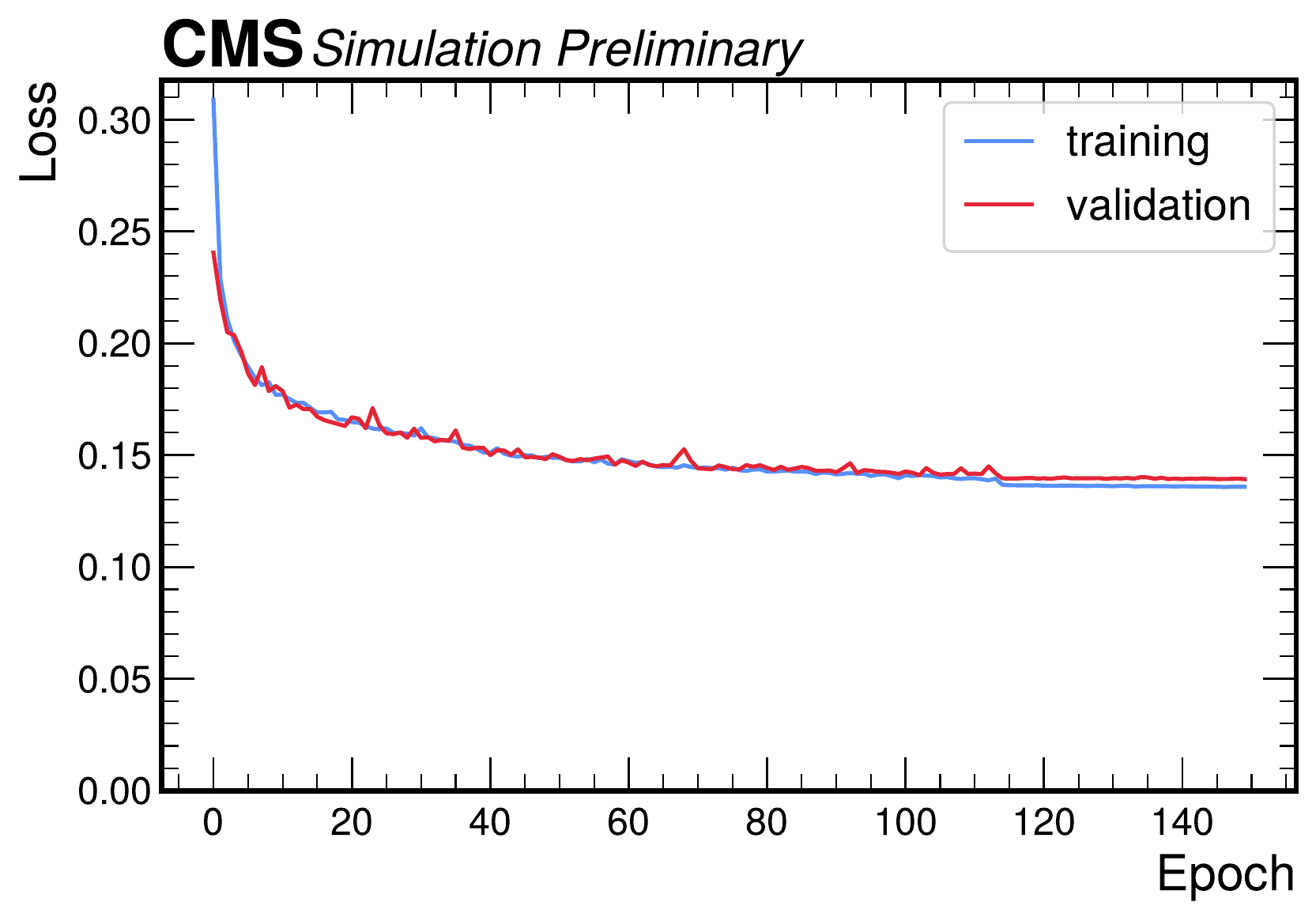}
\caption{Left: the architecture of the CNN, including pre- and post-processing steps. Right: the loss values for the training and validation data.
}
\label{fig:dpcnn}
\end{figure}

The denoising network is a convolutional neural network (CNN) whose architecture is based on Ref.~\cite{Pixar}.
The PyTorch framework~\cite{NEURIPS2019_9015} is used to implement the CNN~\cite{SimDenoising_training}.
The CNN has 9 convolutional layers, each of which creates 100 features using $3\times3$ kernels, shown in Fig.~\ref{fig:dpcnn} (left).
The first layer uses a leaky rectified linear unit (ReLU) activation function with a parameter $\alpha = 0.2$,
because $\Enorm$ can take negative values that would be eliminated by a standard ReLU activation.
It is found that applying the leaky ReLU to the first layer is sufficient to preserve the low-energy fidelity,
so the intermediate layers use the standard ReLU.
No activation function is applied to the last layer, as it performs a regression to predict the final \Enorm value for each pixel.
The transformations are then reversed to obtain the predicted $E$ value.
The CNN output is compared to the high-quality simulation using the mean absolute error, also called the L1 loss.
The training is conducted for 150 epochs with a batch size of 50 images and an initial learning rate of 0.001 that is reduced when the loss value plateaus.
The loss function is also evaluated on the validation dataset, indicating that the CNN was not overtrained, as seen in Fig.~\ref{fig:dpcnn} (right).

\section{Results}

\begin{figure}[htb!]
\centering
\threefig{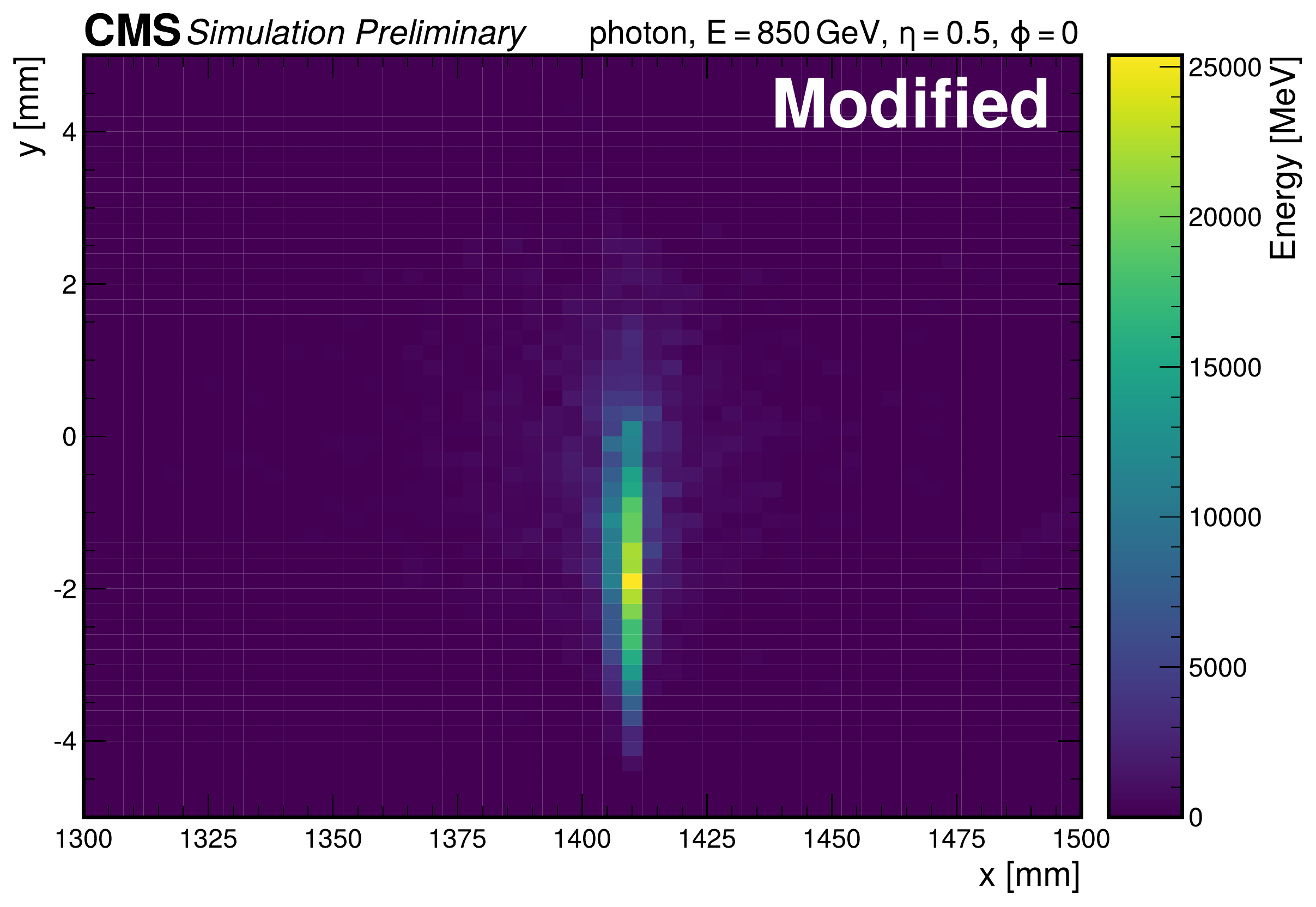}{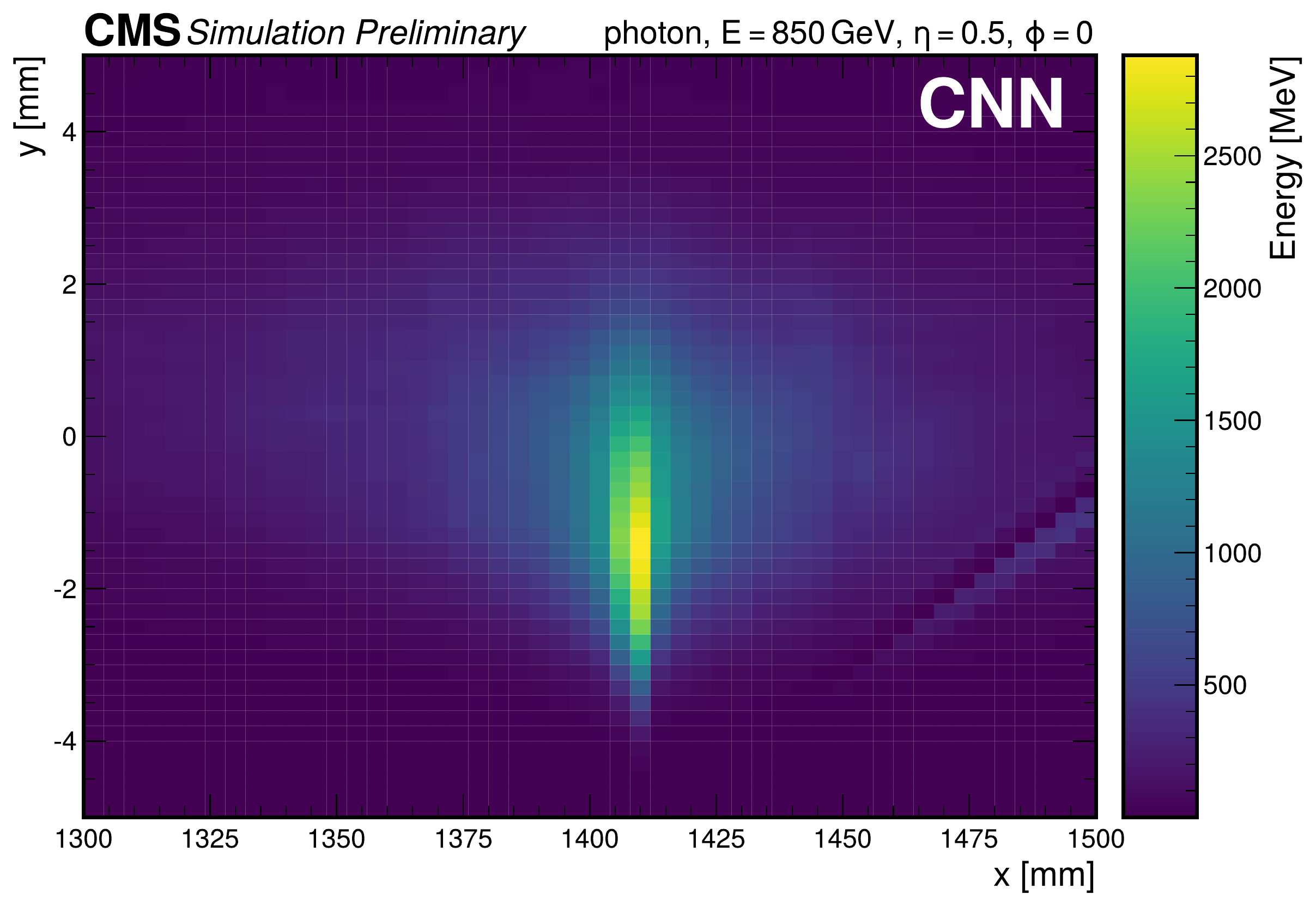}{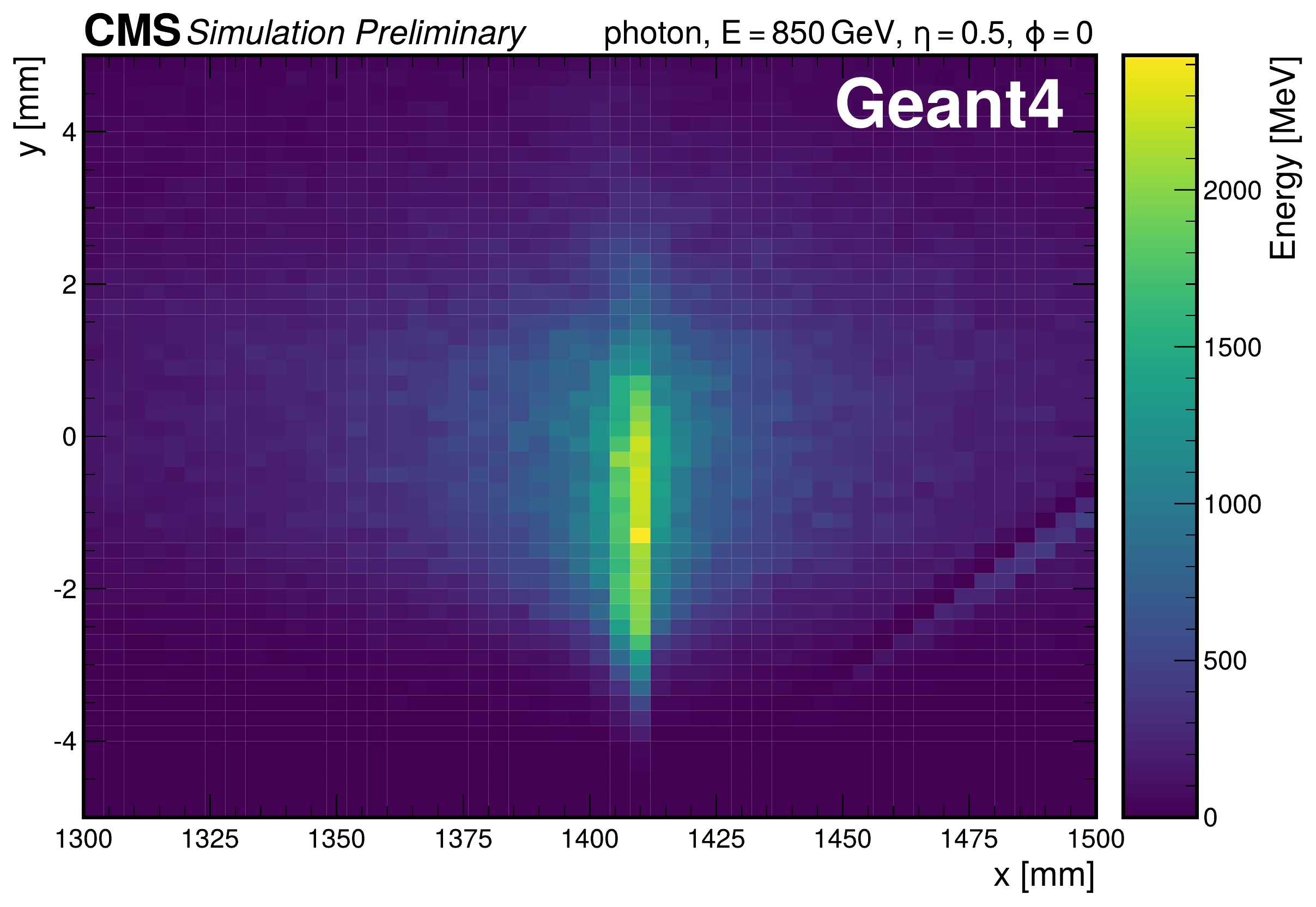}
\caption{An example photon shower event, showing the reduced-quality training input from the modified simulation (left), the CNN output (middle), and \GEANTfour (right).
}
\label{fig:exdenoise}
\end{figure}

\begin{figure}[htb!]
\centering
\threefig{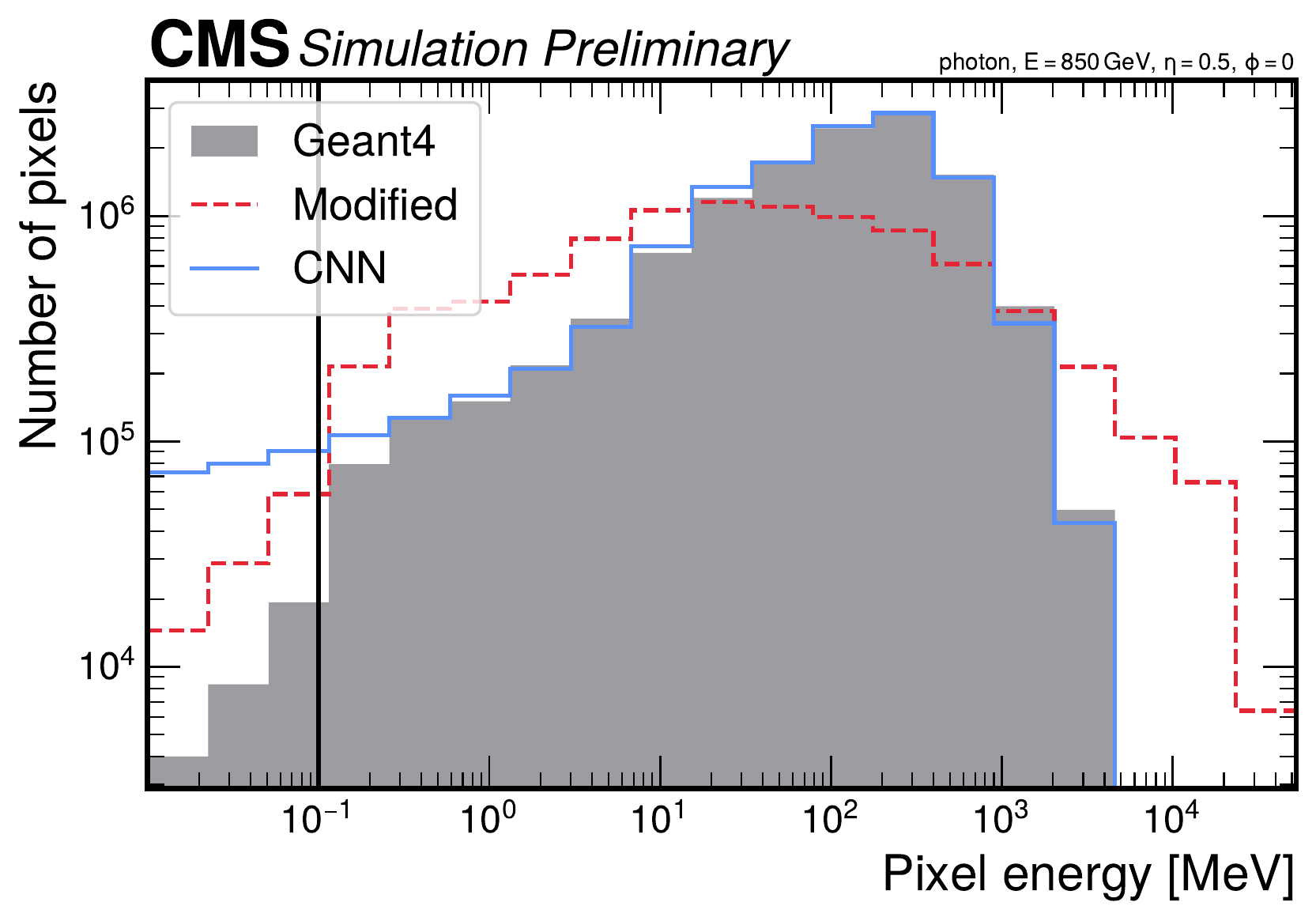}{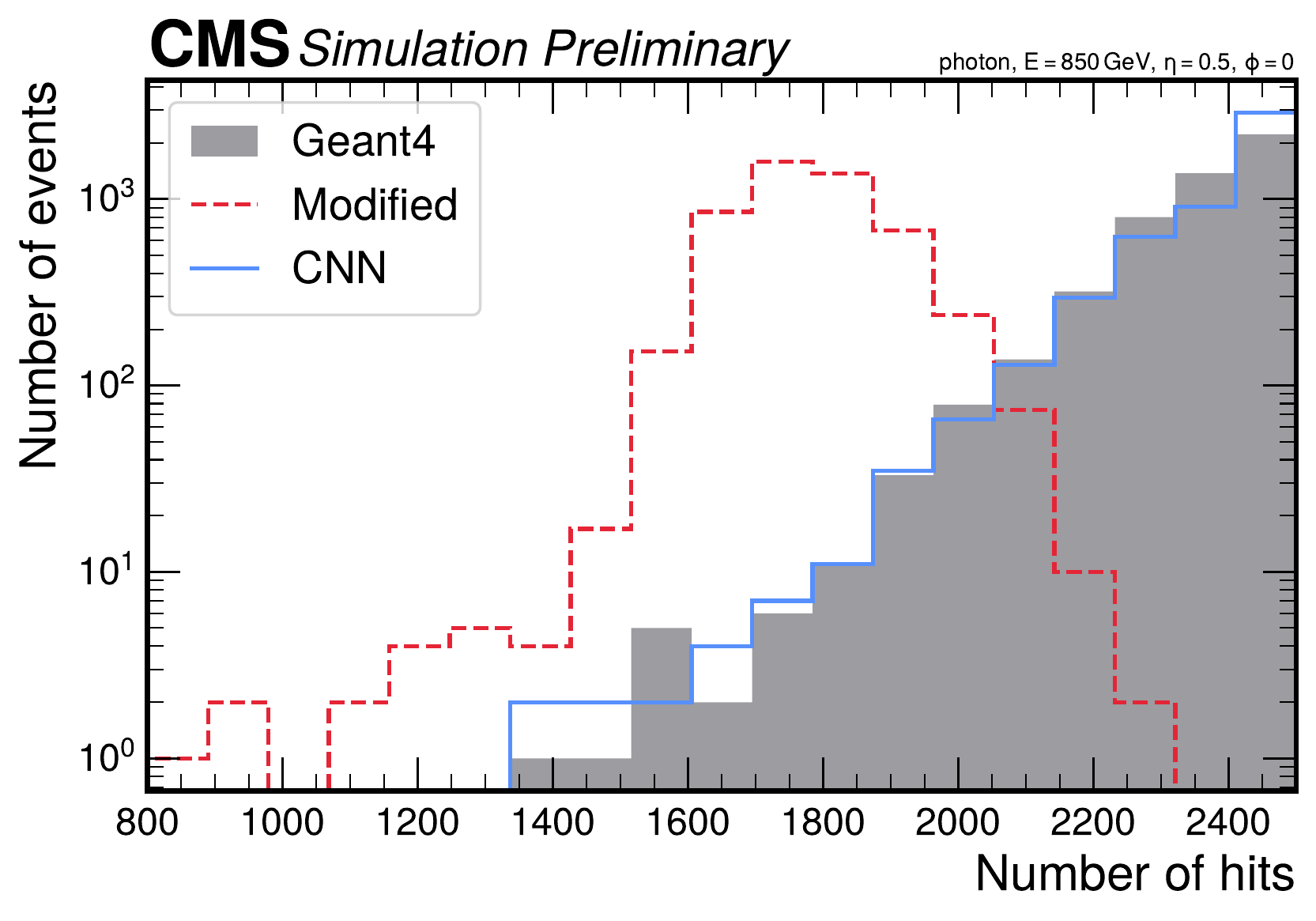}{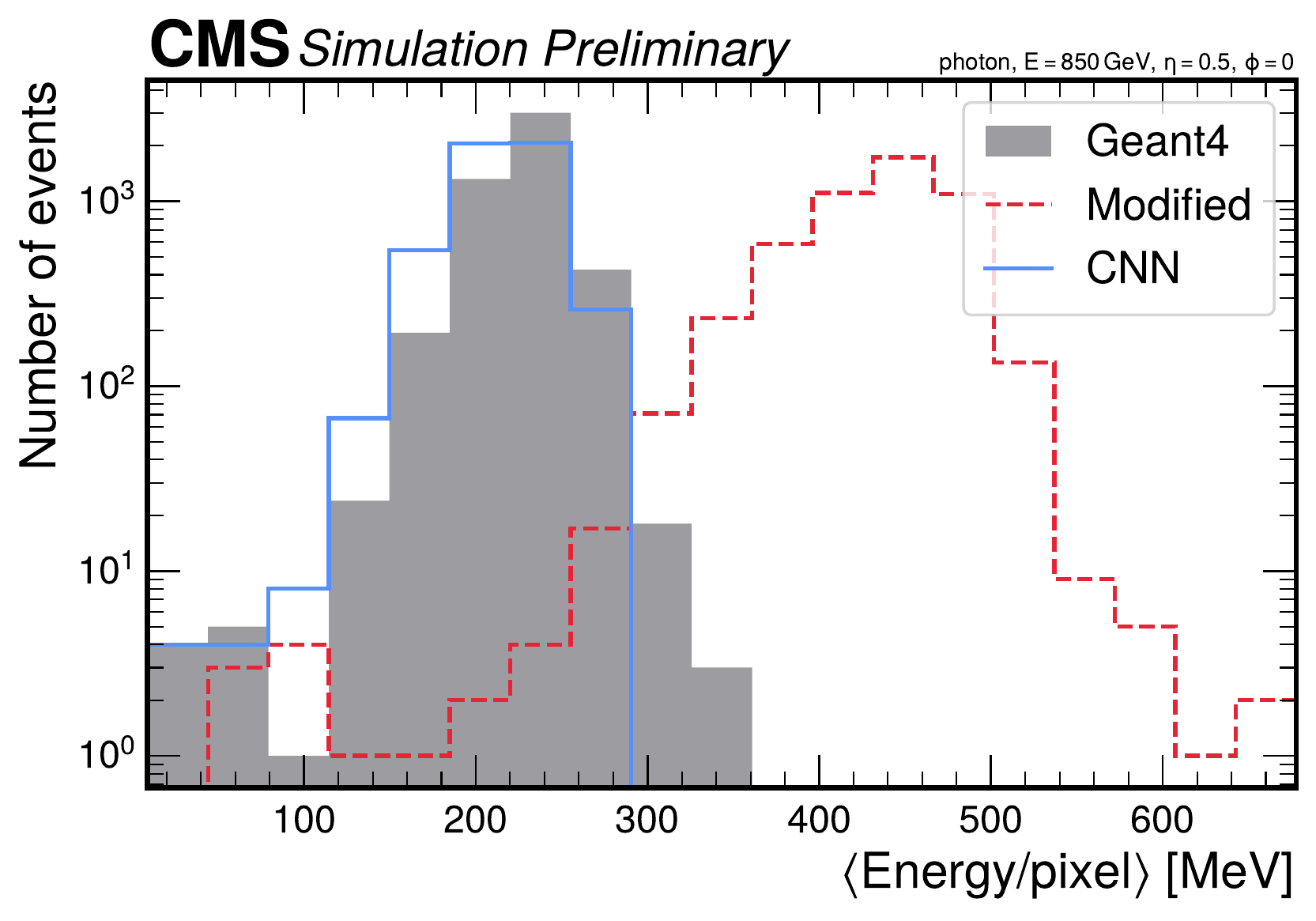}\\
\twofigthree{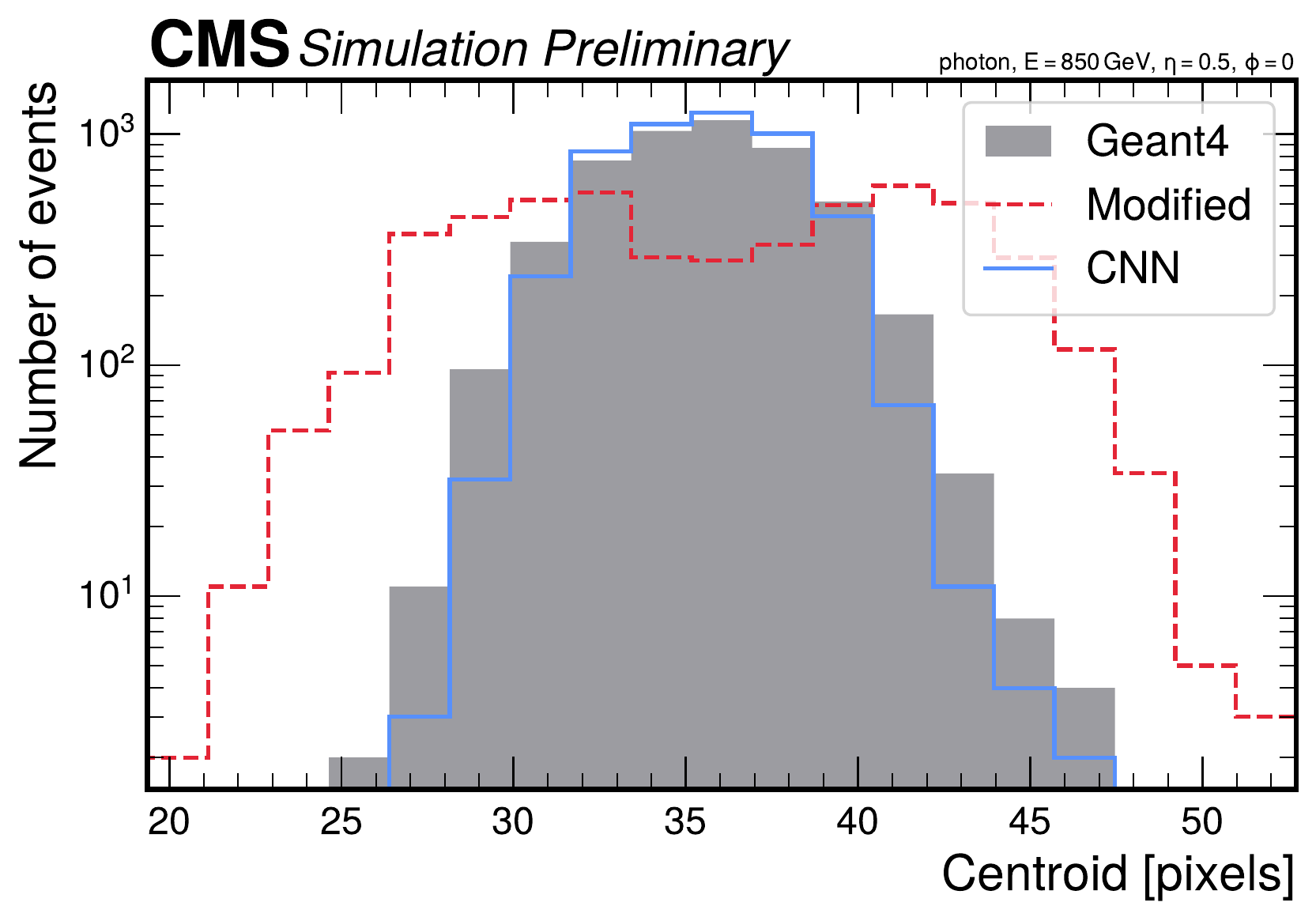}{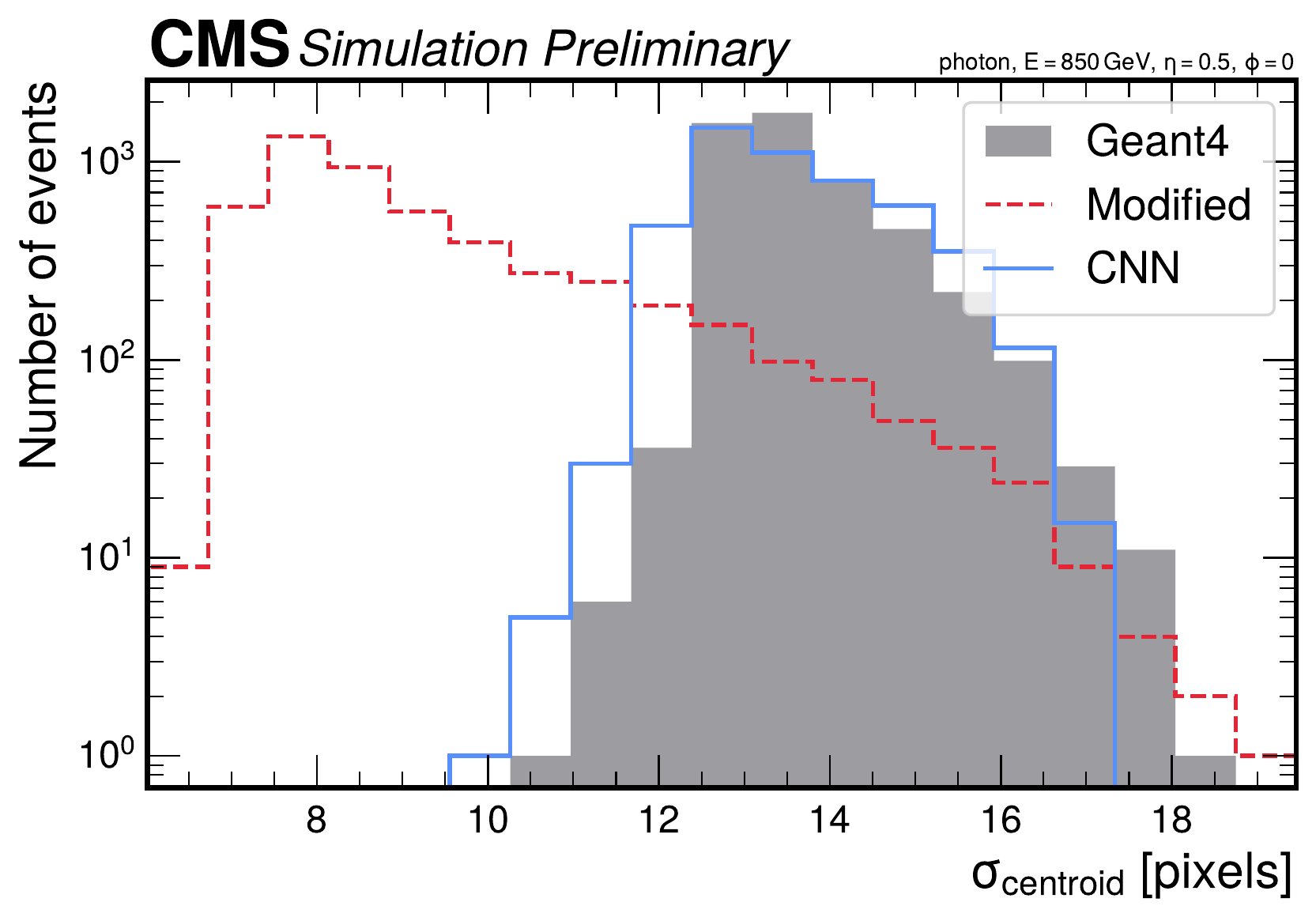}
\caption{Comparisons between the modified simulation, the CNN output, and \GEANTfour for distributions of the per-pixel energy (top left),
the number of hits (top middle), the average energy per pixel (top right), the shower centroid position (bottom left), and the width of the shower centroid (bottom right).
}
\label{fig:qtydenoise}
\end{figure}

Figure~\ref{fig:exdenoise} shows a photon shower event as simulated by the modified simulation, the CNN, and the high-quality simulation.
The CNN output is very similar to the high-quality target, while the lower-quality output from the modified simulation is not.
We proceed to analyze the CNN output by examining both overall distributions and per-event comparisons, using the independent test dataset.
We choose a number of characteristic quantities for particle showers, shown in Fig.~\ref{fig:qtydenoise},
similar to those used in Ref.~\cite{Buhmann:2020pmy} and other studies of ML for detector simulation.
The per-pixel energy \Epixel agrees down to low values; this quantity is difficult for GANs to reproduce accurately without using complicated architectures and extensive post-processing.
The number of hits and average pixel energy are assessed after applying a threshold of $\Epixel > 0.1\MeV$, and they also agree very well with \GEANTfour.
The shower centroid is calculated as the energy-weighted average pixel position,
and its width is computed as the standard deviation of this energy-weighted average.
These quantities show the level of agreement in both the first and second statistical moments of the dataset.

\begin{figure}[htb!]
\centering
\twofigthree{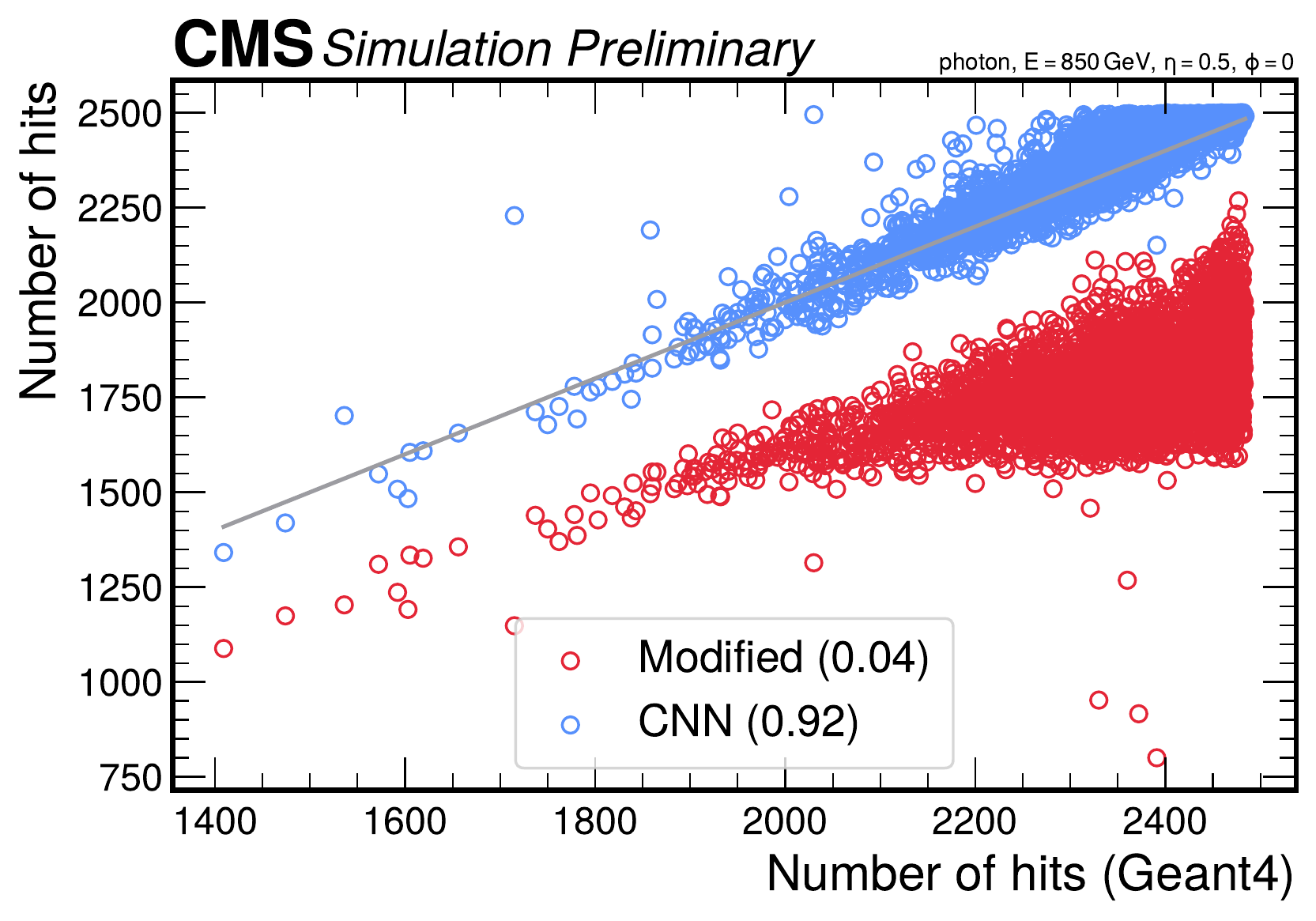}{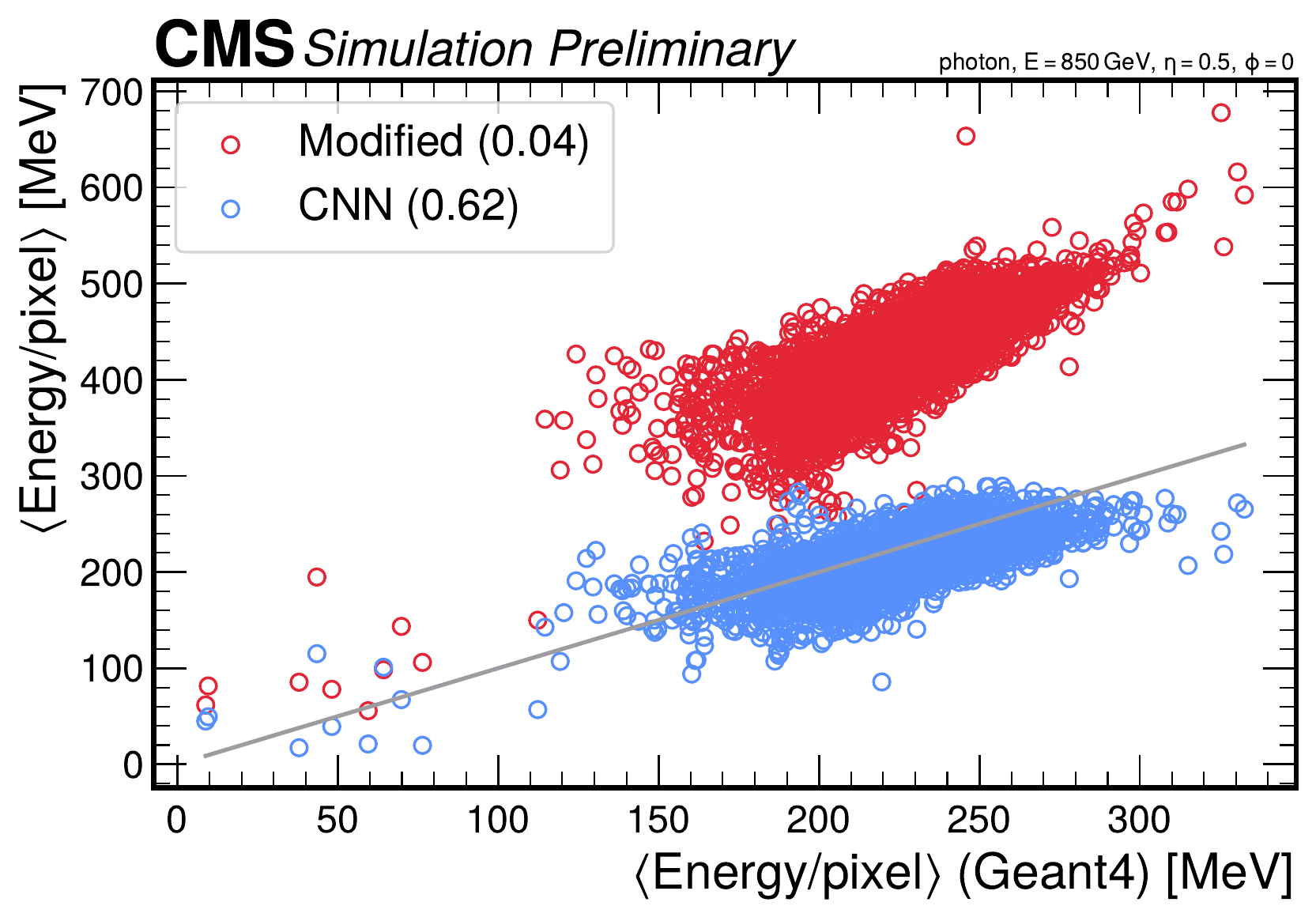}\\
\twofigthree{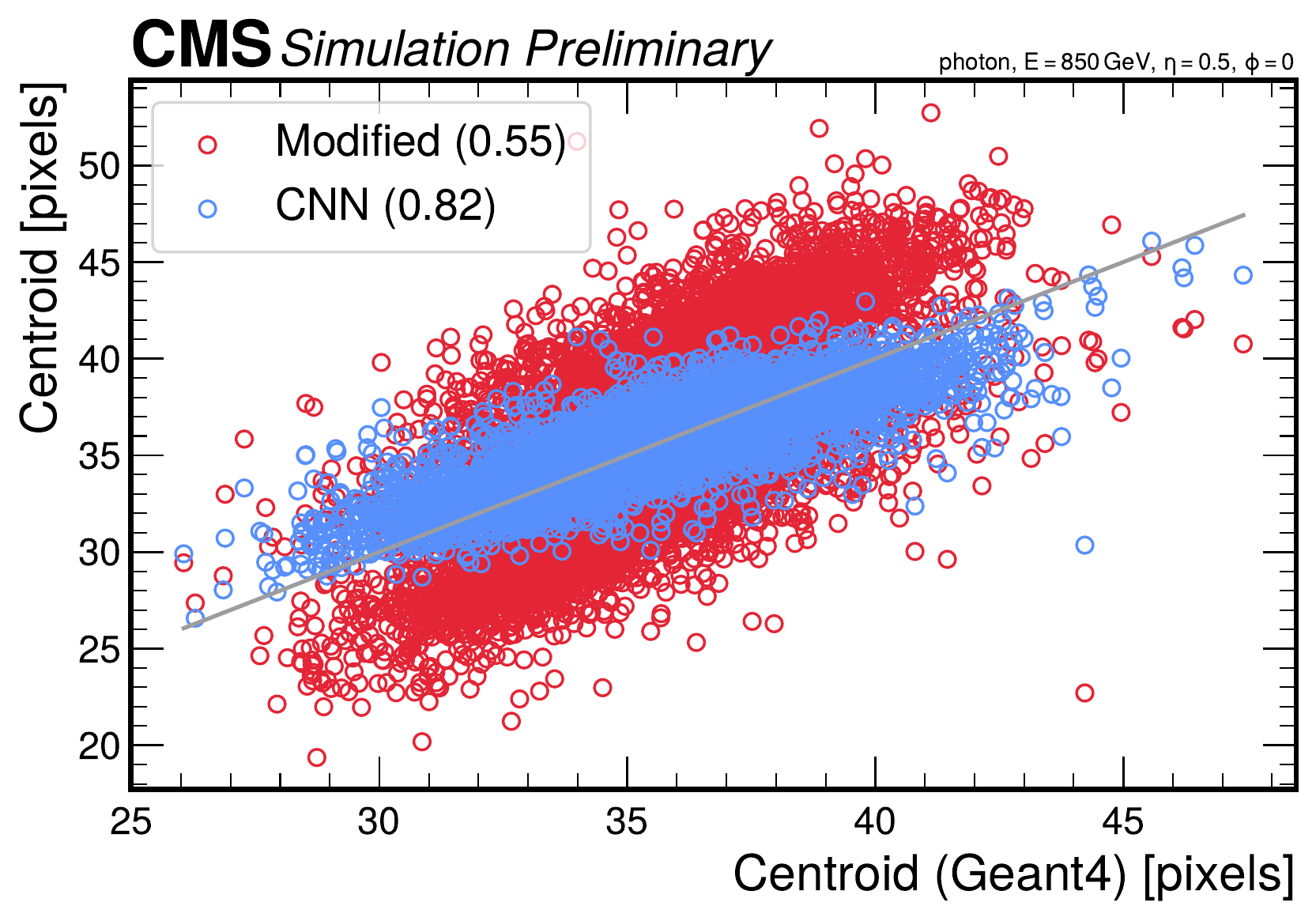}{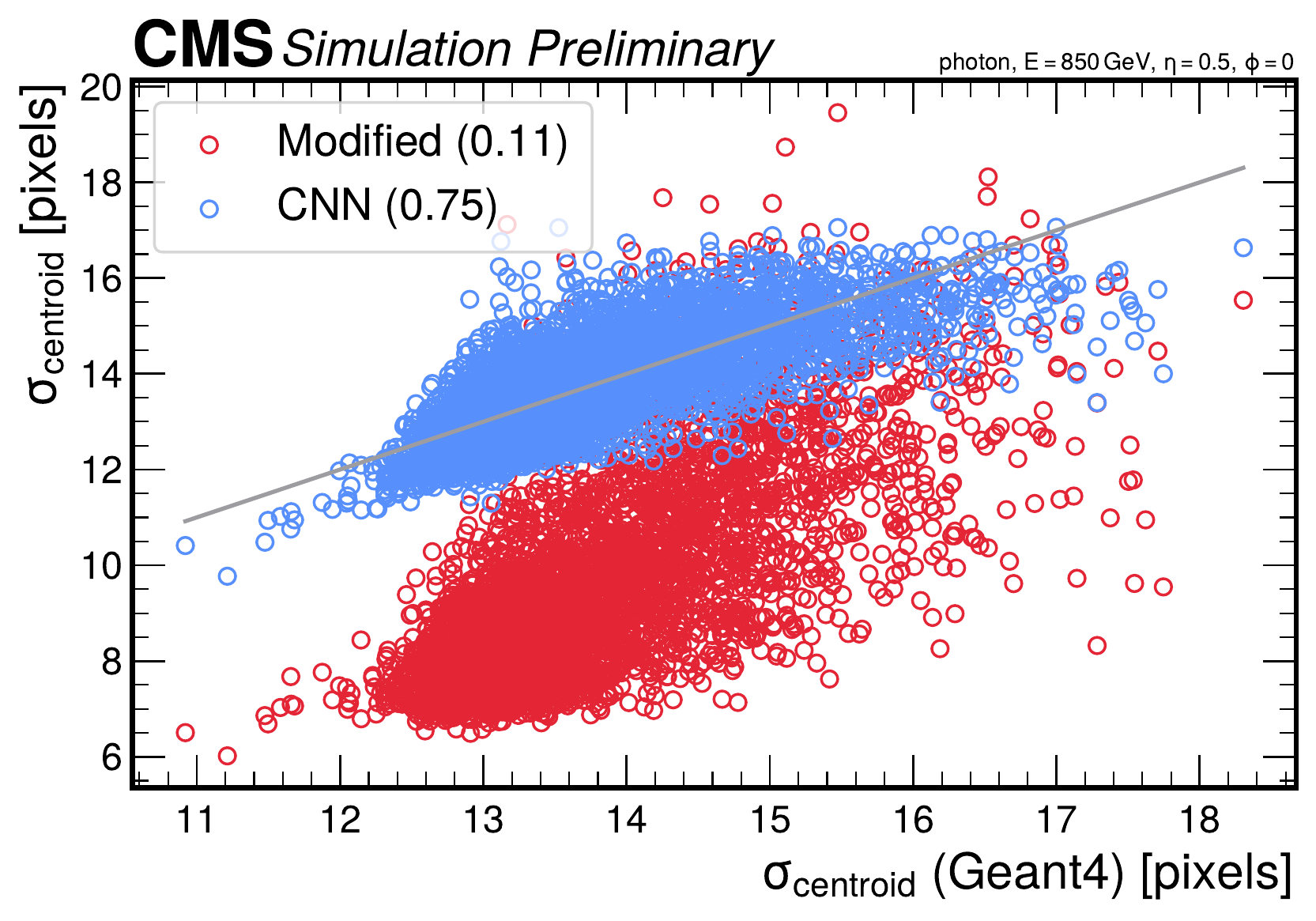}
\caption{Per-image comparisons of the number of hits (top left), the average energy per pixel (top right),
the shower centroid position (bottom left), and the width of the shower centroid (bottom right).
The concordance correlation between \GEANTfour and the others is listed in parentheses,
and the gray line indicates exact agreement.
}
\label{fig:imgdenoise}
\end{figure}

Each pair of low-quality and denoised images has a corresponding high-quality image from the same generated photon event;
this enables the per-image comparisons in Fig.~\ref{fig:imgdenoise}.
Both the low-quality image and the denoised output are compared to \GEANTfour,
and these comparisons are quantified using the concordance correlation~\cite{Concordance} $\concordance = 2\rho \sigma_a \sigma_b/(\sigma_a^2+\sigma_b^2+(\mu_a-\mu_b)^2)$.
This measures the exact agreement between a set of pairs of values $(a,b)$, computed from their correlation coefficient $\rho$ and their means and standard deviations $\mu$ and $\sigma$.
Depending on the quantity, the low-quality simulation has \concordance values of 0.04--0.55, indicating poor agreement with \GEANTfour.
In contrast, the CNN output has substantially higher \concordance values of 0.62--0.92, indicating good agreement.

We also examine the computing performance of the CNN.
For both training and inference, an Nvidia P100 GPU was used.
The training required 2\unit{GB} of memory and took 15 seconds per epoch.
For inference on the GPU, a rate of 800 images per second was achieved without batching, increasing to 8200 images per second when 100 images were batched together.
On a typical CPU, the inference rate was limited to 38 images per second.

\section{Conclusions}

We modified \GEANTfour to act as a fast simulation engine by increasing the production cut to 10\unit{cm}.
This modified simulation was applied to high-energy photon events and its output was compared to the original, high-quality \GEANTfour simulation
in order to train a convolutional neural network (CNN) to perform a regression, predicting the energy value of each pixel in a discretized image formed from the simulation output.
This serves as a proof of concept for a ``denoising'' approach to achieve fast, accurate simulation using machine learning (ML)
in order to address HL-LHC computing challenges.
The strong results are qualitatively competitive with other ML approaches while using a much simpler CNN architecture.
Good agreement is observed in relevant physical quantities for particle showers, examining both overall distributions and per-image correlations.

In the future, we will expand the training data to cover more energy values, particle types, and subdetectors.
Several avenues will be pursued to improve the results even further.
More realistic three-dimensional input images with additional features, beyond just the deposited energy, will be used to train the network.
The network architecture, loss function, and hyperparameters will all be optimized.
Different simulation engines, such as the CMS Fast Simulation application~\cite{Sekmen:2016iql}, will be explored to generate the input data.
Finally, once the approach is mature, the algorithm inference will be implemented in the CMS software.

\section*{References}
\bibliography{cms}
\bibliographystyle{iopart-num}

\end{document}